\documentclass[reprint, showpacs,
 amsmath,amssymb,
 aps,
]{revtex4-1}

\makeatletter
\let\@fnsymbol\@arabic
\makeatother

\usepackage{graphicx}% Include figure files
\usepackage{dcolumn}% Align table columns on decimal point
\usepackage{bm}% bold math
\usepackage{savesym}
\usepackage{amsmath, amsfonts, amsthm, amssymb}
\savesymbol{iint}
\usepackage{txfonts} % this is the font
\restoresymbol{TXF}{iint}
\usepackage[ansinew]{inputenc}
\usepackage{textcomp}
\begin{document}

\title{Low Mass Pseudoscalar Dark Matter in an Extended B - L Model}

\author{E. C. F. S. Fortes and M. D. Tonasse}\thanks{On leave from Campus Experimental de Registro, Universidade Estadual Paulista, Rua Nelson Brihi Badur 430, 11900-000
Registro, SP, Brazil.}
\affiliation{Instituto de F\'\i sica Te\'orica, Universidade Estadual Paulista, Rua Dr. Bento Teobaldo Ferraz 271, 01140-070 S\~ao Paulo, SP, Brazil}

%\altaaffiliation[Permanent address: ]{Campus Experimental de Registro, Universidade Estadual Paulista, Rua Nelson Brihi Badur 430, 11900-000 Registro,SP, Brazil}

\begin{abstract}
We study an extended $B-L$  model, which has in its structure four neutral scalars. In this model, a representative set of parameters enable us to conclude that one of these scalars is a promising candidate for  low-mass dark matter. We introduce an $Z_2$ symmetry, which ensure the stability of the dark matter.  The dominant annihilation process will be through $s$-channel exchange of a scalar   in $b\overline{b}$. So, this is also a Higgs portal dark matter model, but the Higgs decay to dark matter is suppressed  and meets the constraints from invisible decays of Higgs boson. The model is also in agreement with the constraints established by XENON100, CoGeNT and CDMS experiments, maching the relic abundance and the cross section with nucleon.
\end{abstract}

\pacs{12.60.$\pm$i, 95.30.Cq, 95.35.+d, 98.35.Gi, 98.62.Gq}
%12.60.$\pm$i Models beyond the standard model \\
%95.30.Cq Elementary particle processes\\
%95.35.+d Dark matter (stellar, interstellar, galactic, and cosmology) \\
%98.35.Gi Galactic halo \\
%98.62.Gq Galactic halos}

\maketitle

\section{Introduction \label{sec1}}

The researches on dark matter (DM) have evolved rapidly in our days. This is due mainly  to improvements in the techniques of astronomical observations, which have revealed more and better details of the galaxies and galactic clusters structures. Moreover,  a greater sensitivity of underground and spatial experiments has been reached. Recent data about the composition of the Universe are those from the WMAP satellite \cite{Sea03} and the SDSS Collaboration \cite{Tea04}. They have shown that approximately $73 \%$ is dark energy contribution, approximately $4 \%$ is baryonic matter and near $23 \%$ is DM. The DM density is $\Omega h^2 = \rho h^2/\rho_c = 0.1196 \pm 0.0031$ \cite{Planck}, where $\rho_c = 3H_0^2/\left(8\pi G\right)$ is the critical density of the Universe, $H_0$ is the Hubble constant \cite{Bea12}. \par
Nowadays there is a number of experimental events which have some similarity with possible manifestations of DM. Some of them are based on astronomical observations \cite{LH11, HL11}. The others come from the DAMA \cite{Bea10}, CoGeNT \cite{Aea111}, CDMS \cite{Aea211} and XENON \cite{Aea311,Aea411} detectors. These experiments measure the recoil energy of nuclei when it scatters with the DM. DAMA and CoGeNT events rates present a clear annual modulation. They could be interpreted as scattering of DM on the atomic nuclei taking into account the translational motion of the Earth around the Sun.  However, some others detectors as CDMS \cite{Aea211}, XENON10 \cite{Aea311}, and XENON100 \cite{Aea411}, have presented null results related to this modulation. On the other hand, the results of collaborations CoGeNT \cite{KH12} and CDMS \cite{Aea13} are consistent with a component of DM with mass in the range $\left(7 - 10\right)$ GeV and spin independent cross section of $\left(2 - 6\right) \times 10^{-41}$ cm$^2$. Furthermore these results are consistent with the observations of gamma rays from the galactic center \cite{LH11, AK12} and  other regions of the inner Galaxy \cite{HH13}. Recently the LUX Collaboration  announced its first results about light WIMPs limits \cite{NG13}. The  LUX is a more sensitive experiment and it's suitable for detecting low mass dark matter. Again the results of LUX are not compatible with the DAMA, CDMS and CoGeNT for analyzes that take into account isospin conservation. If isospin is violated, the results of LUX disagree with the DAMA and CoGeNT. On the other hand, the results of LUX remain consistent with dark matter mass of 10 GeV, but with the cross-section for proton scattering  two orders of magnitude smaller. \par
As we see, much effort has been carried out in order to make these results compatible among themselves and to be interpreted as DM signals. Despite the difficulties to interpret the present data set as DM evidence, the facts suggest that it  is composed of one or more sort of elementary particles which do not interact or interact very weakly with ordinary matter, except for the gravitational interaction, for which is believed that this particles behave canonically \cite{BE10}. Thus, a good model of electroweak interactions must incorporate a candidate with sufficient properties to play the role of DM. The supersymmetric models present the most popular candidate: the neutralino \cite{EO10}. It is a typical candidate, since it is the lightest supersymmetric particle and the $R$ parity prevents it from interacting with the standard model (SM). The neutralino is an example of what we call {\it cold DM} (CDM), namely a kind of DM which is not relativistic at the time of decoupling between radiation and matter. \par
Besides neutralino, there are other possibilities that include Kaluza-Klein states in models with universal \cite{Kolb, Servant} or warped \cite{Agashe} extra dimensions, stable states in little Higgs theories \cite{Birkedal} and a number of models of extra neutrinos. Other alternative scenarios consider self-interacting DM and warm DM due to the possibility of solving some of the challenges to CDM at the scale of dwarf galaxies without mess up the sucesses of CDM at larger scales \cite{ZV12}.
In order to be consistent with the properties of structure formations, our Universe must be of the $\Lambda$CDM type, i.e., a flat universe with CDM supplemented with a cosmological constant. The models of cold DM of the $\Lambda$CDM type are in excellent agreement with astronomical observations at scales above  few Mpc. On small scales, however, $N$ bodies simulations do not faithfully reproduce the structure of galaxies and clusters of galaxies \cite{SG10, dB09}. \par
The works that propose scalar DM normally extend the SM by introducing a scalar singlet \cite{BP01,BB00}. However, we know that the SM must be extended not only because of the DM problem, but also in order to explain many other problems on its context. Therefore, it becomes interesting to investigate the DM problem in the context of an extended electroweak model which is simple and phenomenologically well motivated. An electroweak model that satisfies these requirements is the $B-L$ one. Therefore, it becomes interesting to investigate the dark matter problem in the context of an extended electroweak model which is simple and phenomenologically well motivated. Here we propose a Higgs pseudoscalar as dark matter, but we work in the context of the B-L model \cite{Appelquist:2002mw}, which has a well-studied phenomenology  \cite{elaine1,basso1}. \par
The original motivation for the $B-L$ model is seeking an explanation for the pattern of neutrino masses and leptonic mixing angles while giving meaning to the accidental $B-L$ symmetry of the SM. In these models, the $B-L$ quantum number, which forbids the neutrino masses in the SM is gauged, yielding an extra neutral gauge boson $Z^\prime$. In order to cancel anomalies, three right-handed neutrinos are added to the model. Thus, the $B-L$ model can provide a small neutrino mass naturally thought the $B-L$ spontaneous symmetry breaking. In order to  break $B-L$, a SU(2)$_L$ complex Higgs singlet is introduced. In this work, to have a DM candidate, we introduce in the model a second complex scalar singlet. This singlet, which obeys a $Z_2$ symmetry, leads to a pseudoscalar which interacts with ordinary matter mainly through the Higgs boson of $125$ GeV, satisfying the requirements to be a candidate for DM. \par

\section{The Model}

The minimal $B-L$ model is developed in greater detail in Ref. \cite{basso1}.  As already mentioned, the SM $B-L$ accidental symmetry is promoted to  an U(1)$_{B-L}$ local symmetry. Thus, this symmetry must be broken, since it does not manifest at low energies. To this task, it is added to SM a SU(2)$_L$ scalar singlet $\chi_1$ with $B-L$ = $2$, which develops a vacuum expectation value (VEV) $x$ and has a vacuum representation as $\chi_1 = x + \xi_1 + i\zeta_1$. The breaking of SU(2)$_L$$\otimes$U(1)$_Y$ to the U(1)$_{\rm Q}$ of electromagnetism remains governed by a SU(2)$_L$ doublet as in the SM, i.e.,
\begin{equation}
\phi = \left(\begin{array}{c}i\omega^+ \\ h^0\end{array}\right).
\label{H}\end{equation}\label{Higgs}
with $B - L = 0$. In the symmetry breaking process, $h^0$ is shifted to $v + \xi + i\zeta$, with $v$ being its VEV. Here, in order to have a scalar DM candidate we add another singlet  also with $B-L = 2$, obeying the reflection symmetry $Z_2: \chi_2 \to -\chi_2$ (the other fields transform trivially under $Z_2$), which is shifted as $\chi_2 = \xi_2 + i\zeta_2$, conserving $Z_{2}$. The Higgs potential is given  by
\begin{eqnarray}
V\left(\phi, \chi_1, \chi_2\right) && = \mu^2\Phi^\dagger \Phi + \mu_1^2\chi_1^*\chi_1 + \mu_2^2\chi_2^*\chi_2^2 + \lambda_1\left(\Phi^\dagger\Phi\right)^2 + \cr
&& + \lambda_2\left(\chi_1^*\chi_1\right)^2 + \lambda_3\left(\chi_2^*\chi_2\right)^2 + \lambda_4\left[\left(\chi_1^*\chi_2\right)^2 + {\mbox{H. c.}}\right] + \cr
&& +\lambda_5\chi_1^*\chi_2\chi_2^*\chi_1 + \lambda_6\left(\chi_1^*\chi_2 + {\mbox{H. c.}}\right)^2 + \cr
&& + \Phi^\dagger\Phi\left(\lambda_7\chi_1^*\chi_1 + \lambda_8\chi_2^*\chi_2\right).
\label{V}\end{eqnarray}
In the potential (\ref{V}), the constants $\mu$, $\mu_1$ and $\mu_2$ have dimension of mass, while $\lambda_{1, \ldots, 8}$ are dimensionless. From the real part of the potential (\ref{V}) we get the square masses
\begin{subequations}\begin{eqnarray}
m_1^2 & = & \mu_2^2 + \lambda_8v^2 + \left(2\lambda_4 + \lambda_5 + 4\lambda_6\right)x^2; \\
m_2^2 & = & 2\left[\lambda_1v^2 + \lambda_2x^2 - \sqrt{\left(\lambda_1v^2 + \lambda_2x^2\right)^2 + \lambda_7^2v^2x^2}\right] \\
m_3^2 & = & 2\left[\lambda_1v^2 + \lambda_2x^2 + \sqrt{\left(\lambda_1v^2 + \lambda_2x^2\right)^2 + \lambda_7^2v^2x^2}\right].
\end{eqnarray}
From the imaginary sector we have a Higgs boson with mass
\begin{equation}
m_4^2 = \mu_2^2 + \lambda_8v^2 + \left(-2\lambda_4 + \lambda_5\right)x^2
\end{equation}\label{m12}\end{subequations}
and two Gosdstone which are eaten by the neutral gauge bosons $Z$ and $Z^\prime$. \par
The respective eigenstates are
\begin{subequations}\begin{align}
h_1 & = \xi, \quad h_2 & = c_\xi\xi_1 + s_\xi\xi_2, \quad h_3 & = -s_\xi\xi_1 + c_\xi\xi_2, \\
h_4 & = \zeta, \quad G_1 & = \zeta_1, \quad G_2 & = \zeta_2,
\end{align}\end{subequations}
where $h_i$ $\left(i = 1, \ldots, 4\right)$, $G_1$ and $G_2$ are physical eigenstates with $G_1$ and $G_2$ being the Goldstone bosons. The mixing $c^2_\xi = 1 - s_\xi^2 = \cos^2\theta_\xi$ is defined by
\begin{equation}
\theta_\xi = \arctan{\frac{\lambda_7vx}{\lambda_2x^2 - \lambda_1v^2 - \sqrt{\left(\lambda_2x^2 - \lambda_1v^2\right)^2 + \lambda_7^2v^2x^2}}}.
\end{equation}
In Table \ref{tab1} and \ref{tab2} we present relevant couplings for the scalars of the model are presented in Table \ref{tab1}. Therefore, the Eqs (\ref{m12}) and the Table \ref{tab1} suggest that $h_{1}$ is the Higgs boson of 125 GeV. The scalar we propose as DM candidate is $h_4$. Hence, it is important that $m_{h_4} \approx 10$ GeV, that is, in the region suggested by most of the experiments (see Sec. \ref{sec1}). \par
From the gauge sector, the masses of the gauge bosons $Z$ and $Z^\prime$ are given by:
\begin{equation}
m_Z^2 = \frac{1}{4}\left(g^2 + g_1^2\right)v^2, \qquad m^2_{Z^\prime} = {g_1^\prime}^2x^2,
\end{equation}
where $g$, $g_1$ and $g^\prime_1$  is the coupling constants of SU(2), U(1)$_Y$ and U(1)$_{B-L}$ groups, respectivelly. The limits imposed by LEP and Tevatron establish that $m_{Z^{\prime}}/g_1^{\prime}\geq 7$ TeV \cite{cac} and the LEP experiments also provided a lower bound of $x \geq 3.5$ TeV \cite{cac}. \par
The lepton Yukawa interactions are given by
\begin{equation}
-{\cal L} = y^{e}_{jk}\overline{\ell^\prime_{jL}}e^\prime_{kR}\phi  + y^{\nu}_{jk}\overline{\ell_{jL}}N^\prime_{kR}\sigma_2\phi^* + y^{M}_{jk}\overline{\left(N^\prime_R\right)^c_j}N^\prime_{kR}\chi_1 + {\mbox{h. c.}},
\label{L}\end{equation}
where $\ell_{iL} = \left(\begin{array}{cc} \nu_i & e_i \end{array}\right)_L^T$, $N_{iR}$ are heavy neutrinos, $y^e_{jk}$, $y^\nu_{jk}$, $y^M_{jk}$ are Yukawa constants and $i, j, k$ are family indexes. We are not assuming inter-families mixing for the neutrinos, so that we assume $y_{jk}^e = y_e$, $y_{jk}^\nu = y_\nu$ and $y_{jk}^M = y_M$ \cite{basso1}.  Therefore, from the Lagrangian (\ref{L}) we obtain the neutrino masses
{\footnotesize
\begin{center}
\begin{table}[h]
%\begin{center}
\caption{\footnotesize\baselineskip = 12pt Trilinear Higgs interactions.
\label{tab1}}
\begin{ruledtabular}
\begin{tabular}{l|l}
\hline
   $h_{1}h_{1}h_{1}$ & $4\lambda_{1}v$ \\
 $h_{1}h_{1}h_{2}$ & $2\lambda_7c_\xi x$ \\
   $h_{1}h_{1}h_{3}$ & $2\lambda_7s_\xi x$ \\
   $h_{1}h_{2}h_{2}$ & $2\left(\lambda_7c_\xi^2 + \lambda_{8}s_\xi^2\right)v$ \\
  $h_{1}h_{2}h_{3}$ & $4\left(\lambda_7 - \lambda_8\right)c_\xi s_\xi v$ \\
  $h_{1}h_{3}h_{3}$ & $2\left(\lambda_7s_\xi^2 + \lambda_8c_\xi^2\right)v$ \\
  $h_{1}h_{4}h_{4}$ & $2\lambda_8v$ \\
  $h_{2}h_{2}h_{2}$ & $2\left[2\lambda_2c_\xi^2 + \left(2\lambda_4 + \lambda_5 + 4\lambda_6\right)s_\xi^2\right]c_\xi x$ \\
  $h_{2}h_{2}h_{3}$ & $2\left[2\left(3\lambda_2 - 2\lambda_4 - \lambda_5 - 4\lambda_6\right)c_\xi^2 + \left(2\lambda_4 + \lambda_5 + 4\lambda_6\right)s_\xi^2\right]s_\xi x$ \\
  $h_{2}h_{3}h_{3}$ & $2\left[\left(2\lambda_4 + \lambda_5 + 4\lambda_6\right)c_\xi^2 + 2\left(3\lambda_2 - 2\lambda_4 - \lambda_5 - 4\lambda_6\right)s_\xi^2\right]c_\xi x$ \\
  $h_{2}h_{4}h_{4}$ & $-2\left(2\lambda_4 - \lambda_5\right)c_\xi x$ \\
  $h_{3}h_{3}h_{3}$ & $2\left[\left(2\lambda_4 + \lambda_5 + 4\lambda_6\right)c_\xi^2 + 2\lambda_2s_\xi^2\right]s_\xi x$ \\
   $h_{3}h_{4}h_{4}$ & $-2\left(2\lambda_4 - \lambda_5\right)s_\xi x $\\
\hline
\end{tabular}
\end{ruledtabular}
%\end{center}
\end{table}
\end{center}
}
\begin{subequations}\begin{eqnarray}
m_{\nu_i} & = & \frac{1}{2}\left(y_Mx_1 - \sqrt{y_\nu^2v^2 + y_M^2x_1^2}\right), \\
m_{N_i} & = & \frac{1}{2}\left(y_Mx_1 + \sqrt{y_\nu^2v^2 + y_M^2x_1^2}\right)
\end{eqnarray}\end{subequations}
with the eigenstates
{\footnotesize
\begin{center}
\begin{table}[h]
%\begin{center}
\caption{\footnotesize\baselineskip = 12pt Neutrino couplings to scalars.
\label{tab2}}
\begin{ruledtabular}
\begin{tabular}{l|lll}
\hline
      & $\overline\nu_\ell\nu_\ell$ & $\overline\nu_\ell\nu_h$ & $\overline\nu_h\nu_h$ \\
\hline
$h_1$ & $2y_\nu c_\nu s_\nu$ & $-y_\nu\left(c_\nu^2 - s_\nu^2\right)$ & $-2y_\nu c_\nu s_\nu$ \\
$h_2$ & $-2y_Ms_\nu^2c_\xi$ & $2y_Mc_\nu s_\nu s_\xi$ & $2y_Mc_\nu^2c_\xi$ \\
$h_3$ & $2y_Ms_\nu^2s_\xi$ & $-2y_Mc_\nu s_\nu s_\xi$ & $2y_Mc_\nu^2s_\xi$ \\
\hline
\end{tabular}
%\end{center}
\end{ruledtabular}
\end{table}
\end{center}
}
\begin{equation}
\nu_i = c_\nu\nu_{\ell_i} + s_\nu\nu_{h_i}, \qquad N_i = -s_\nu\nu_{\ell_i} + c_\nu\nu_{h_i},
\end{equation}
where $c_\nu = \cos\theta_\nu$ and $s_\nu = \sin\theta_\nu$ and $\tan^2\theta_\nu = -m_{N_i}/m_{\nu_i}$. \par

\section{Dark Matter Abundance}

To study the evolution of the numerical density $n$ of $h_{4}$, the DM candidate, at the temperature $T$ in the early Universe the Boltzmann equation can be written in simplified form
\begin{equation}
\frac{dY}{dy} = -\sqrt{\frac{\pi
g_*}{45G}}\frac{m_{h_4}}{y^2}\langle\sigma_{\rm ann}|\textrm{v}|\rangle\left(Y^2 -
Y_{eq}^2\right),
\label{bol}\end{equation}
where $Y = n/s$, $s$ is the entropy per unity of volume, $Y_{eq}$ is the $Y$ value in the thermal equilibrium,  $y = m_{h_4}/T$. The parameter $G$ is the universal constant of gravitation, $\sigma_{\rm ann}$ is the cross section for annihilation of the particle $h_4$ and $\textrm{v}$
is the relative velocity. In Eq. (\ref{bol}), the symbol $\langle\rangle$
represents thermal average. The term $g_*$ is a
parameter that measures the effective number of degrees of freedom at freeze-out, which is expressed as
\begin{equation}\label{aa}
g_{*}=\sum_{i=bosons}g_{i}\left(\frac{T_{i}}{T}\right)^{4} +\frac{7}{8}\sum_{i=fermions}g_{i}\left(\frac{T_{i}}{T}\right)^{4}.
\end{equation}
Note that  $g_*$ is a function of $T$, and the sums in Eq. (\ref{aa}) runs over only those species with mass $m_{i}\ll T$ \cite{KT98}.  Considering $T \gtrsim 300$ GeV, in this model $g_*=113$ and, because of the assignment of values to the parameters in Sec. \ref{sec4}, the particles included in this calculation will be all the species of standard model plus 1 extra Higgs and 3 right-handed neutrinos. In this case, the $Z^{\prime}$ and the other two scalars, which are also contained in the model, didn't appear in the calculation of $g_*$ since they are heavier than $300$ GeV. \par
To find $Y_0$, the present value of $Y$,  Eq. (\ref{bol}) must be integrated between $y = 0$ and $y_0 = m_{h_4}/T_0$. Once this value is found, the contribution of $h_4$ to DM density is
\begin{equation}
\Omega_{h_4} = \frac{m_{h_4} s_0Y_0}{\rho_c}.
\end{equation}
Considering the model studied here and the parameters set, the annihilation cross section
mediated by $h_{2}$ is dominant over the one mediated by $h_{1}$. The DM annihilates mainly in $b\overline{b}$, $c\overline{c}$ and $\tau^{+}\tau^{-}$.  The cross section for annihilation of $h_4$ into fermions $f$ is given by
\begin{eqnarray}\label{sigma}
\sigma(h_{4}h_{4}\rightarrow f\overline{f}) & = & \frac{N_{c}}{64\pi s}\sqrt{\frac{s-4m_{f}^{2}}{s-4m_{h_{4}}^{2}}}\frac{g_{244}^{2}g^{2}s_{\xi}^{2} m_{f}^{2}}{m_{W}^{2}}\times \cr
&& \times\frac{s-4m_{f}^{2}}{(s-m_{h_{2}}^{2})^{2}+m_{h_{2}}^{2}\Gamma_{h_{2}}^{2}}
\end{eqnarray}
where $N_{c}$ denotes de color number, $g$ is coupling constant of SU(2)$_L$, $m_{f}$ is the fermion mass and $g_{244}$ is the strenght of the interaction $h_{2}h_{4}h_{4}$ (Table \ref{tab1}). \par
If there are no resonances and coannihilations, it's
required that the thermal average annihilation cross section is
$\langle\sigma_{\rm ann}|\textrm{v}|\rangle \sim 3\times 10^{-26}$ cm$^3$/s at
the temperature of freeze-out ($T_f\simeq m_{h_4}/x_f$) with $x_{f}\simeq
$ 20 to 30 in order to have a relic abundance $\Omega_{h_4} =0.1196 \pm
0.0031$. \par
In this Letter we use the MicrOMEGAs package, which employs the
Runge-Kutta method to solve numerically the Boltzmann equation (\ref{bol})
\cite{Belanger:2010pz}. All the interactions of the minimal $B-L$ model are
given in Ref. \cite{basso1}. We had implemented all the interactions in
the CalcHEP package \cite{Belayev} and in MicrOMEGAs. \par
{\begin{figure}[ht]
 \begin{center}
  \includegraphics[width=5 cm, height=4 cm]{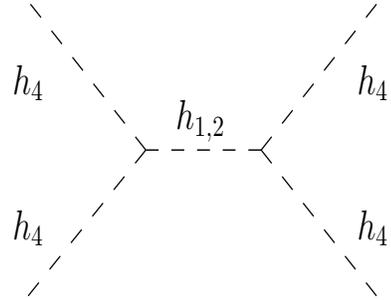}
   \caption{Feynman diagrams involved in the process $h_4h_4 \rightarrow
h_4h_4$.\label{diagramas}}
\end{center}
\end{figure}}
In Fig. \ref{diagramas} we presents the Feynmann diagrams which contribute
to scaterring of $h_4$.

\section{Results and Comments \label{sec4}}

In this section we present the parameter choice for the model. In order to have $h_4$ as DM candidate we had used the following inputs: $\mu_{2}=3170.19$,  $\lambda_{1}=2.3$, $\lambda_{2}=4\times 10^{-2}$, $\lambda_{3}=1$, $\lambda_{4}=-9.9\times 10^{-2}$, $\lambda_{5}=-6\times 10^{-1}$, $\lambda_{6}=9.916\times 10^{-2}$, $\lambda_{7}=3\times 10^{-4}$, $\lambda_{8}=10^{-7}$, $v = 246$ GeV and $x=5000$ GeV. This particular choice will result in the following masses for the scalars: $m_{h_{1}}=126.8$ GeV, $m_{h_{2}}=746.15$ GeV, $m_{h_{3}}=2000$ GeV and the DM candidate will have a mass $m_{h_{4}}=10.2$ GeV. For the neutrino sector we have chosen $y_M=10^{-6}$ and $y_\nu= 10^{-8}$, so the masses for the light and heavy neutrino will be respectivelly $m_{\nu_{1}}=m_{\nu_{2}}=m_{\nu_{3}}\simeq 3\times 10^{-10}$ GeV and $m_{N_{1}}=m_{N_{2}}=m_{N_{3}}=5\times 10^{-3}$ GeV. \par
So, this parameter choice leads to  $\sigma |\textrm{v}|=2.63\times 10^{-26}$ cm$^{3}$/s, $\Omega =0.11$ and the dominant annihilation channels for $h_{4}$ will be in fermions, with 87\% in $b\overline{b}$, 7\% in $\tau\overline{\tau}$ and 6\% in $c\overline{c}$. In addition, our spin-independent elastic cross sections are close to $\sigma_{I}\simeq (2-5)\times 10^{-41}$ cm$^{2}$ established by XENON100 contraints for low-mass DM \cite{Aea411}. We had obtained $\sigma_{I,p} = 1.17\times 10^{-41}$ cm$^{2}$ and $\sigma_{I,n} = 1.19\times 10^{-41}$ cm$^{2}$ to collisions with the proton and neutron, respectively. It's interesting to notice that there are other parameters regions that can also lead to the right experimental results. \par
Our DM candidate  is light. Global fits  put limits on the resulting invisible decay of the Higgs boson such that $B\left(H \rightarrow {\rm invisible}\right) < 0.19 (0.38)$ at $95\%$ CL \cite{Greljo,Belanger2013}.  But in the scenario chosen here, the Higgs decay into the DM is suppressed. The coupling of Higgs to DM is presented in Table \ref{tab1} and depends essentially in the parameters $\lambda_{8}$ and $v$. The Higgs coupling to DM should be smaller than the SM bottom Yukawa coupling, considering our parameter choice the branching will be $B\left(H \rightarrow {\rm DM}\right)=2.9\times 10^{-8}$, which is very safe.

\section{Conclusion}

We have proposed a scenario where $B-L$ model has a potential low mass DM candidate. The model has four scalar bosons, two which are heavy, one which play the role of the Higgs with mass of $125$ GeV and the other which plays the role of DM candidate.  The DM candidate is a pseudoscalar and the dominant annihilation processes are via scalars exchange in $s-$channel. \par
The model has a interesting motivation and it's a Higgs portal to DM without spoiling the present constraints for invisible Higgs decays. Besides, the spin-independent elastic cross section is in good agreement with the results of experiments CoGeNT and CDMS discussed in Sec. \ref{sec1} for a DM with mass in the order of $(7 - 10)$ GeV \cite{Aea411}. On the other hand, the LUX experiment, presented results that are not fully compatible with last ones. With our parameter choice the results of GoGeNT and CDMS are reproduced. The results of LUX also can be reproduced with another set of parameters of the model. As the experiments are in tension among themselves, it's not possible to reach all the results at the same time. \par
 We had shown here that the minimal B - L model, with the addition of a singlet complex scalar is a theoretically self-consistent model for dark matter. So, the model is a good candidate to be reached in the future experiments of LHC and direct DM seaches. Differently of the  Higgs of 125 GeV, the other scalar $h_{2}$, which dominates the annihilation processes has its decay predominantly in invisible DM decays.

\acknowledgments
One of us, E. C. F. S. F., thanks the Funda\c c\~ao de Amparo \`a Pesquisa do Estado de S\~ao Paulo (FAPESP) for the financial support (Process N$^{\d o}$ 2011/21945-8). M. D. T. thanks the Instituto de F\'\i sica Te\'orica of the UNESP for hospitality.

\end{document}